%% file: main.tex
\documentclass[sigconf]{acmart}
\usepackage{cleveref}

\renewcommand\footnotetextcopyrightpermission[1]{} 
\usepackage{multirow}
\usepackage{enumitem}
\usepackage{mleftright}
\usepackage{placeins}
\usepackage{tcolorbox}
\tcbuselibrary{listings}

\setcopyright{acmlicensed}
\copyrightyear{2018}
\acmYear{2018}
\acmDOI{XXXXXXX.XXXXXXX}
\acmConference[Conference acronym 'XX]{Make sure to enter the correct
  conference title from your rights confirmation email}{June 03--05,
  2018}{Woodstock, NY}
\acmISBN{978-1-4503-XXXX-X/18/06}

\begin{document}
\title{Equity vs. Equality: Optimizing Ranking Fairness for Tailored Provider Needs}

\author{Yiteng Tu}
\affiliation{%
    \institution{Tsinghua University}
    \city{Beijing}
    \country{China}
}
\email{tyt24@mails.tsinghua.edu.cn}

\author{Weihang Su}
\affiliation{%
    \institution{Tsinghua University}
    \city{Beijing}
    \country{China}
}

\author{Shuguang Han}
\affiliation{%
    \institution{Alibaba}
    \city{Beijing}
    \country{China}
}

\author{Yiqun Liu}
\affiliation{%
    \institution{Tsinghua University}
    \city{Beijing}
    \country{China}
}

\author{Qingyao Ai*}
\affiliation{%
    \institution{Tsinghua University}
    \city{Beijing}
    \country{China}
}

\input{secs/0abs}


\maketitle

\input{secs/1intro}
\input{secs/2related}
\input{secs/3frame}

\input{secs/4method}
\input{secs/5experi}

\input{secs/6result}
\input{secs/7conclu}


\bibliographystyle{ACM-Reference-Format}
\bibliography{sample-base}

\end{document}

%% file: secs/0abs.tex
\begin{abstract}
Ranking plays a central role in connecting users and providers in Information Retrieval (IR) systems, making provider-side fairness an important challenge.
While recent research has begun to address fairness in ranking, most existing approaches adopt an equality-based perspective, aiming to ensure that providers with similar content receive similar exposure. 
However, it overlooks the diverse needs of real-world providers, whose utility from ranking may depend not only on exposure but also on outcomes like sales or engagement. 
Consequently, exposure-based fairness may not accurately capture the true utility perceived by different providers with varying priorities.
To this end, we introduce an equity-oriented fairness framework that explicitly models each provider’s preferences over key outcomes such as exposure and sales, thus evaluating whether a ranking algorithm can fulfill these individualized goals while maintaining overall fairness across providers. 
Based on this framework, we develop EquityRank, a gradient-based algorithm that jointly optimizes user-side effectiveness and provider-side equity. 
Extensive offline and online simulations demonstrate that EquityRank offers improved trade-offs between effectiveness and fairness and adapts to heterogeneous provider needs.
\end{abstract}

%% file: secs/1intro.tex
\section{Introduction}
Ranking plays a vital role in Information Retrieval (IR) systems like search engines~\cite{croft2010search, langville2006google, macdonald2013whens} and recommender systems~\cite{chen2019top, karatzoglou2013learning, yang2022effective} which connect users and content providers. 
From the user’s perspective, effective ranking algorithms ensure that the most relevant and useful information appears at the top, helping users quickly find what they need and improving their overall experience~\cite{ma2018modeling, papenmeier2023undr}.
From the provider’s perspective, ranking determines the visibility of their content, which directly affects the amount of traffic and the income they receive from the platform~\cite{yang2022effective,zehlike2020reducing}.
Due to its central role in shaping information access and distribution, ranking not only shapes individual experiences but also drives broader economic and social outcomes~\cite{singh2018fairness}.
Given its dual impact on users and providers, recent research has increasingly focused on balancing ranking effectiveness with fairness for content providers~\cite{biega2018equity, morik2020controlling, singh2019policy, yang2021maximizing}.

A common approach to provider fairness centers on exposure-based metrics, under the assumption that content providers benefit proportionally from the visibility their items receive. 
For instance, amortized exposure fairness~\cite{biega2018equity, singh2018fairness} aims to ensure that providers offering similar content receive similar levels of exposure over time. 
While this equality-based perspective is intuitive and easy to implement, it fails to capture the diversity of provider objectives in practice. 
Different providers may target different user segments, promote heterogeneous content types, or pursue varying business goals, all of which shape their desired outcomes. 
As a result, identical exposure may not translate into comparable benefits across providers.
From a sociological perspective, this distinction is often captured by the contrast between \textbf{equality}—treating everyone the same—and \textbf{equity}, which emphasizes allocating resources based on specific needs~\cite{matsumoto1996culture,xu2023p}. 
Motivated by this distinction, we explore how \textbf{equity}-oriented fairness can be better aligned with real-world provider diversity in IR systems.

To make equity-oriented fairness practical, it is important to consider the tailored needs and goals of different content providers. 
In real-world scenarios, providers may have varying expectations depending on the type of content, promotional strategies, or the business stage they are in~\cite{yang2020search}. 
In addition, the benefits that providers gain from IR systems are often influenced by various factors beyond exposure, such as revenue, retention, user satisfaction, and user engagement~\cite{patro2022fair}.
For example, well-established and widely recognized brands and merchants often prioritize direct profit and higher conversion rates (CVR) over exposure. 
However, for most emerging brands and entrepreneurs, while also valuing revenue, they are typically more eager to gain visibility and exposure as soon as possible. 
A similar dynamic also exists within the same brand between classic, long-standing products and newly launched ones.
These differences mean that a one-size-fits-all view of fairness, based solely on exposure, is often too simplistic. 
However, most existing work assumes a uniform relationship between exposure and provider utility, which may result in fairness measures that misrepresent what providers truly care about.
This limitation calls for new fairness definitions and methods that can better capture and adapt to the diverse objectives of content providers.

To address the limitations of existing fairness paradigms, we propose a novel framework that explicitly accounts for provider-specific needs. 
Motivated by empirical observations that content providers may prioritize different performance metrics, such as click-through rates (CTR) and CVR, especially in e-commerce scenarios~\cite{yang2020search}, we focus on a representative setting where provider utility primarily derives from two fundamental outcomes: item exposure and sales (or user examinations and purchases)~\cite{solomon2020consumer,vakratsas1999advertising}. 
Under this formulation, we assume that each provider has tailored preferences over these two factors and introduce an \textbf{equity-oriented fairness metric} that evaluates whether a ranking algorithm can fulfill these heterogeneous needs while maintaining overall fairness across providers.

Our proposed metric offers greater flexibility and generality: it can be naturally extended to incorporate additional utility dimensions or simplified to recover amortized exposure fairness as a special case. 
Building on this foundation, we further introduce a gain-based ranking algorithm, \textbf{EquityRank}, which jointly optimizes user-side ranking effectiveness and provider-side fairness from an equity perspective. 
The algorithm is implemented as a gradient-based optimization method guided by our fairness metric, and we evaluate it through both offline simulations (with known relevance labels) and online simulations (where relevance must be inferred dynamically during the ranking process). 
Experimental results on multiple large-scale recommender datasets demonstrate that EquityRank better aligns with the personalized goals of users and providers, leading to improved trade-offs between ranking effectiveness and fairness.

In conclusion, our contributions are threefold: (1) We introduce a novel equity-oriented ranking framework that considers the diverse needs of providers. (2) We propose a fairness-aware ranking algorithm, EquityRank, based on this framework. (3) We conduct extensive experiments to show that EquityRank can achieve superior performance in both offline and online settings. It provides ranking results that better align with providers' tailored needs compared to existing state-of-the-art methods.

%% file: secs/2related.tex
\section{Related Work}
\label{sec:related}

\subsection{Fairness in Ranking}
\label{subsec:def}
Due to the profound impact of ranking on content providers~\cite{singh2018fairness, zehlike2017fa, morik2020controlling}, provider fairness in ranking has received considerable attention in the IR community~\cite{bigdeli2022gender, naghiaei2022cpfair, raj2022measuring, ge2022explainable}. 
Existing work on ranking fairness presents diverse definitions, but it is generally agreed that they can be broadly categorized into two types: probability-based fairness and exposure-based fairness~\cite{patro2022fair,zehlike2021fairness}. 
Probability-based fairness typically involves ensuring that the number or proportion of items from protected groups (e.g., age, gender, race) in the top-ranked positions should not fall below a certain value~\cite{celis2017ranking, zehlike2017fa, geyik2019fairness}. 
On the other hand, exposure-based fairness operates under the premise that the platform's total exposure resources are limited~\cite{patro2022fair}. 
It assigns exposure levels to each position in the rank list based on expected user attention or click probability and aims to distribute exposure as fairly as possible among different items (individual fairness) or groups (group fairness), for example, by ensuring that exposure is proportional to relevance~\cite{singh2018fairness, singh2019policy, biega2018equity}. 
Additionally, similar to the definition of exposure fairness, \citet{morik2020controlling} and \citet{singh2018fairness} propose a variation called impact fairness, which requires that the number of clicks is proportional to the relevance.
Compared to probability-based fairness, exposure-based fairness aligns better with the principle of group fairness~\cite{morik2020controlling, patro2022fair}, and has therefore garnered more attention from researchers. 
However, despite their difference in settings, both assume that providers benefit equally from each item's exposure in rankings and do not consider their personalized needs.
Therefore, they only focus on the equality perspective of fairness and ignore the diversity of provider needs in real-world scenarios.

\subsection{Exposure Fairness Algorithms}
As discussed previously, existing studies on provider fairness algorithms are mostly derived from exposure fairness.
The concept of exposure fairness (or amortized fairness) is initially introduced by \citet{singh2018fairness} and ~\citet{biega2018equity}, who present methods utilizing linear programming or 0-1 integer programming to address it. 
However, the high time complexity of these methods poses challenges for their application on large-scale datasets. 
FairCo~\cite{morik2020controlling} offers a dynamic learning-to-rank (LTR) algorithm that integrates unbiased estimators for relevance and fairness and dynamically adjusts rankings. 
Drawing inspiration from the MMR~\cite{carbonell1998use} method for search result diversification, MMF~\cite{yang2021maximizing} employs a greedy selection approach to maximize both marginal relevance and fairness in constructing rankings. 
PG-Rank~\cite{singh2019policy} and PL-Rank~\cite{oosterhuis2021computationally}  also address the balance between relevance and fairness in the LTR process. They utilize the Plackett-Luce model~\cite{plackett1975analysis} to model the ranking strategy and offer valuable insights into the Plackett-Luce model.
Furthermore, MCFair~\cite{yang2023marginal} introduces uncertainty as an additional factor beyond relevance and fairness, while TFROM~\cite{wu2021tfrom} incorporates the concept of customer fairness alongside provider fairness, aiming for equitable utility distribution among users.
Nonetheless, to the best of our knowledge, all existing ranking algorithms for provider fairness focus on the optimization of equality-oriented objectives.
How to account for the diverse needs of providers in fairness optimization is still an open question.

%% file: secs/3frame.tex
\input{tabs/notation}

\section{Preliminaries}
\label{sec:problem}

\subsection{Service Workflow} \label{subsec:rank_service}
Let there be $m$ product providers $\mathcal{G} = \{g_1, g_2, ..., g_m\}$ on the ranking platform. 
These providers can be regarded as various item groups like different industries, companies, brands, etc. 
Each provider $g$, offers $n_g$ items represented as $\mathcal{I}_{g} = \{\tau_{g}^{1}, \tau_{g}^{2}, ..., \tau_{g}^{n_g}\}$. 
Let $\mathcal{I} = \bigcup_{g \in \mathcal{G}}\{\mathcal{I}_{g}\}$ be all items available on the platform, with the total count of items being $n$.
Then, given a ranking request, the goal of a ranking algorithm is to rank the items in $\mathcal{I}$ and present a sub-list of them to optimize the system goal.

Let $T$ be a specific time step when a user $u_T$ issues a ranking request to the system.
Here we assume that only one user can issue a request at each time step, but this formulation can easily be extended to scenarios where multiple user sessions happen simultaneously.
After the system receives the request, the relevance estimator of the ranking algorithm predicts the relevance of each candidate item to the user, and then the system ranks these items based on their estimated user relevance, recommending the top-$K$ items to the user. 
Formally, let the result ranking for user $u_T$ be $\pi_T$, which can be viewed as a permutation of the item set $\mathcal{I}$ with the top-$K$ items: $\pi_T \in \rm{Permute}(\mathcal{I})[:K]$. 
Then, based on $\pi_T$, the user would interact with the items accordingly, such as browsing, clicking, purchasing, etc. 
These interactions could lead to direct benefits to the item providers and also serve as important signals for the system to estimate the user-item relevance.
Specifically, we only consider two types of interactions in this paper for simplicity, i.e., the user's examination of a specific item (or the exposure of an item, denoted as $\mathbf{e}$), and the user's purchase of an item after examining it (or the sale of an item, denoted as $\mathbf{b}$). 
They are considered the two most crucial and fundamental factors in marketing for content providers, directly tied to their profits~\cite{solomon2020consumer,vakratsas1999advertising}.
A summary of the notations is in Table~\ref{tab:notations}.

\subsection{Interaction Hypothesis}\label{subsubsec:gain}
To understand and simulate user interactions with the ranking system, we adopt the following interaction hypothesis widely used in existing ranking literature~\cite{yang2023marginal,yang2021maximizing}.
First, we assume that a user will only purchase an item if it is shown to him/her (exposed) and meets his/her needs (relevant):
\begin{equation}
\mathbf{b} = \begin{cases}\mathbf{r}, & \text { if }  \mathbf{e} = 1, \\ 0, & \text { otherwise}, \end{cases}
\end{equation}
where $\mathbf{b}, \mathbf{r}, \mathbf{e}$ are binary random variables representing whether an item is bought by the user, relevant to the user, and exposed to the user, respectively. 
Similar to ~\citet{chuklin2022click}, we model the probability of purchasing an item as:
\begin{equation}
P(\mathbf{b} = 1) = P(\mathbf{r} = 1)P(\mathbf{e} = 1),
\end{equation}
where $P(\mathbf{r} = 1)$ indicates the probability of relevance of the item to the user (denoted as $r(*)$ for convenience in the subsequent sections). 
$P(\mathbf{e} = 1)$ stands for user's examination probability. 

Following previous studies~\cite{oosterhuis2021unifying, yang2023marginal}, we assume that there are two types of biases when users interact with the ranking results, namely position bias and selection bias:
\begin{itemize}[leftmargin=*]
\item \textbf{Position Bias}~\cite{craswell2008experimental}: Users are more likely to examine top-ranked items than those on lower ranks, the examination probability is solely decided by the item's ranking position, denoting as $p_{\rm{rank}(\tau | \pi)}$, where $\rm{rank}(\tau | \pi)$ is item $\tau$'s position in rank list $\pi$.
\item \textbf{Selection Bias}~\cite{ovaisi2020correcting}: Users cannot see items not presented to them. For convenience, we assume that users cannot examine items ranked beyond the $K$-th position in the rank list.
\end{itemize}
Based on the above assumptions, users' examination probability can be formulated as:
\begin{equation}\label{eq:examprob}
P(\mathbf{e} = 1 | \tau, \pi) = \begin{cases}p_{\rm{rank}(\tau | \pi)}, & \text { if }  \rm{rank}(\tau | \pi) \leq K ,\\ 0, & \text { otherwise. } \end{cases}
\end{equation}

\subsection{User (Ranking) Effectiveness}\label{subsec:user_effect}
As the primary goal of IR systems is to retrieve results that fulfill users' information needs, user utility is usually considered as one of the most important perspectives for ranking evaluation.
It is also referred to as the ranking effectiveness in the existing literature~\cite{yang2023marginal}. 
Assuming that users receive benefits from and only from interacting with items relevant to their needs, then, based on the interaction hypothesis above, we can directly measure user utility based on the relevant items in a rank list.
A widely used metric to quantify this is Discounted Cumulative Gain (DCG)~\cite{jarvelin2002cumulated}. 
For a given user $u$ and rank list $\pi$, the DCG value for the top-$k_c$ results is:
\begin{equation}
\mathrm{DCG}@k_c(\pi) = \sum_{k=1}^{k_c} r(\pi[k], u) \cdot p_k,
\end{equation}
where $\pi[k]$ is the $k$-th item in the rank list, $r(*)$ is the relevance function, and $p_k$ denotes the examination probability of the $k$-th position in the rank list, which decreases as the position $k$ increases~\cite{singh2018fairness, yang2023marginal}.
For convenience, we will abbreviate the probability of relevance between the user and the item $r(\pi[k], u)$ as $r(\pi[k])$.

Once DCG has been defined, we can introduce the concept of normalized DCG~(NDCG). It normalizes the metric within the range of $[0, 1]$ by utilizing the DCG value of the ideal ranking $\pi^*$ where items are arranged in descending order of relevance.
Specifically, in offline settings where all users and relevance scores are recorded in advance, we adopt average NDCG~(aNDCG) to measure effectiveness. 
In contrast, in online settings where the system is receiving real-time user requests on the fly, we can measure the user utility with cumulative NDCG~(cNDCG)~\cite{jarvelin2002cumulated}:
\begin{equation}\begin{aligned}\label{eq:acndcg} 
\mathrm{eff.} = \begin{cases} 
\mathrm{aNDCG}@k_c = \frac{1}{|\mathcal{U}|}\sum_{u}\frac{\mathrm{DCG}@k_c(\pi_u)}{\mathrm{DCG}@k_c(\pi^*_u)}, & \text{if} \ \text{offline}, \\
\mathrm{cNDCG}@k_c(T) = \sum_{t=1}^T\gamma^{T-t}\frac{\mathrm{DCG}@k_c(\pi_t)}{\mathrm{DCG}@k_c(\pi^*_t)}, & \text{if} \ \text{online}. \\
\end{cases}
\end{aligned}
\end{equation}

While user effectiveness is important in an IR system, previous studies have found that greedily optimizing ranking systems for user utility in each session may lead to unhealthy environments that promote monopoly, which could hurt the platform and user experience in the long term~\cite{morik2020controlling,singh2018fairness}.
Therefore, ranking utility from the content provider side, i.e., the provider fairness, is also considered to be an important objective for ranking optimization.
In contrast to previous studies that focus on provider fairness from an equality perspective, in this paper, we propose an equity-oriented provider fairness framework that considers the tailored needs of different content providers.

\section{Equity-oriented Fairness} \label{sec:measure}
In this section, we introduce our proposed concept of equity-oriented fairness. 
Specifically, it consists of two parts: provider needs modeling and the definition of fairness based on provider gains.

\subsection{Provider Needs}
Typically, providers benefit directly from a ranking platform through user interactions on their items (e.g., products, articles, services, etc.). 
However, depending on their competition environment and marketing strategies, different providers may value different types of user interactions differently~\cite{yang2020search}.
For simplicity, as discussed in~\S\ref{subsec:rank_service}, we only take the two most critical types of user interactions into account, i.e., examination and purchase.
Specifically, to model the differences between each provider's needs for user examination and purchase, we define the gains of a provider $g$ obtained from a user's examination and purchase with $v_e^g$ and $v_b^g$, respectively. 
The values of $v_e^g$ and $v_b^g$ not only represent the characters of the providers' items but also reflect their preferences over different factors in ranking. 
They can be collected from the ranking platform and the content providers in advance (e.g. from the price and production cost of each item, from the promotional fee that a provider is willing to pay for an item's exposure, purchase, etc.), and can be used to model the diverse needs of the providers.
Consequently, the gain (and its expectation) harvested by a provider $g$ from a rank list $\pi$ associated with a user $u$, $G(g | \pi, u)$,  can be calculated as\footnote{It is important to note that, for simplicity, we assume that all items under the same provider have identical values for the same category of the gain (i.e., $v_e$ or $v_b$). Actually, our framework and methodology can seamlessly adapt to situations where different items result in varying gains, simply by replacing the group-wise $v_e, v_b$ with the item-wise $v_e^\tau, v_b^\tau$, which we leave for future studies.}:
\begin{equation}
G(g | \pi, u) = \sum_{\tau \in \mathcal{I}_{g} \cap \pi } P(\mathbf{b} = 1 | \tau, \pi) v_b^g + P(\mathbf{e} = 1 | \tau, \pi) v_e^g,
\end{equation}
\begin{equation}\begin{aligned}\label{eq:gain_pertime}
\mathbb{E}(G(g | \pi, u)) & = \mathbb{E} \sum_{\tau \in \mathcal{I}_{g} \cap \pi } P(\mathbf{b} = 1 | \tau, \pi) v_b^g + P(\mathbf{e} = 1 | \tau, \pi) v_e^g \\  
& =  \sum_{\tau \in \mathcal{I}_{g} \cap \pi } \mathbb{E}\left[P(\mathbf{b} = 1 | \tau, \pi) v_b^g + P(\mathbf{e} = 1 | \tau, \pi) v_e^g\right] \\ 
& = \sum_{\tau \in \mathcal{I}_{g} \cap \pi } P(\mathbf{e} = 1 | \tau, \pi) (r(\tau, u) \cdot v_b^g + v_e^g) \\ 
& = \sum_{\tau \in \mathcal{I}_{g} \cap \pi} p_{\rm{rank}(\tau | \pi)} (r(\tau, u) \cdot v_b^g + v_e^g).
\end{aligned}\end{equation}

\subsection{Gain-based Provider Utility and Fairness}
Based on the needs of each provider, we now introduce a general fairness metric based on their gains.
Without the loss of generality, we assume that each provider $g$ has an expected gain value or expected gain ratio $y^g$ to be achieved.\footnote{Ideally, $y^g$ should be determined through negotiation between the platform and the providers to represent the tailored needs of different providers.
For instance, it could ensure that the expected revenue of certain disadvantaged providers is not too low or that merchants paying higher promotional fees can expect greater returns.
Since we primarily focus on introducing a new framework rather than the pricing mechanism in this paper, and due to the unavailability of real-world e-commerce data, we do not delve into the details of how it is obtained. We only consider the relative magnitude.}
Then, to achieve fairness over providers with different expected gains with a limited bandwidth of user traffic, the ranking systems should help the providers to receive gains that align proportionally with their expected values. 
Formally, $\forall g_i, g_j \in \mathcal{G}$, we should have $\frac{G(g_i)}{y^{g_i}} = \frac{G(g_j)}{y^{g_j}}$. 
Specifically, following the measurement in \cite{oosterhuis2021computationally} to avoid the zero denominator problem, we assess the disparity in average gains between pairs of providers and define the provider-side (un)fairness metric (i.e., $\mathrm{fair.}$ and $\mathrm{unfair.}$) as:
\begin{equation}
\label{eq:fair}
\mathrm{fair.}(T) = -\mathrm{unfair.}(T),
\end{equation}
\begin{equation}\label{eq:unfair}
\mathrm{unfair.}(T) = \frac{1}{m(m-1)} \sum_{i=1}^m\sum_{j=1}^m\left(G(g_i, T)y^{g_j} - G(g_j, T)y^{g_i}\right)^2,
\end{equation}
\begin{equation}\label{eq:gain}
G(g_i, T) = \frac{1}{T} \sum_{t=1}^T G(g_i |\tau_t, u_t ).
\end{equation}

\subsection{Generalizability}
Our proposed gain-based provider fairness framework offers better generalizability as it can achieve both equity and equality. 
First, it can seamlessly accommodate the diverse needs of content providers on multiple factors in ranking by computing the gain of each provider based on their unique weighting of each factor (e.g., including but not limited to user examination and purchase in this paper).
This provides the possibility to define and measure provider fairness from the equity perspective, i.e., whether the systems can provide resources corresponding to each provider's need for success. 

Second, existing exposure fairness and impact fairness~\cite{singh2018fairness, morik2020controlling} can be seen as a special case of our gain-based fairness metric.
Take exposure fairness as an example, in existing studies, the exposure of an item and a group is defined as:
\begin{equation}
\label{eq:exposure}
E(\tau, T) = \sum_{t=1}^T\sum_{k=1}^{K} p_k \mathbb{I}[\pi_t[k] == \tau],
\end{equation}
\begin{equation}
E(g, T) = \frac{1}{n_{g}} \sum_{\tau \in \mathcal{I}_g} E(\tau, T),
\end{equation}
where $\mathbb{I}$ denotes the indicator function and $p_k$ is the examination probability.
After defining exposure, the idea of exposure fairness is that each group's exposure should be proportionate to its merit (i.e., relevance) to users:
\begin{equation}
\frac{E(g_i, T)}{r(g_i)} = \frac{E(g_j, T)}{r(g_j)}, \forall g_i, g_j \in \mathcal{G},
\end{equation}
\begin{equation}
\label{eq:grouprel}
r(g_i) = \frac{1}{n_{g_i}}\sum_{\tau \in \mathcal{I}_{g_i}} r(\tau).
\end{equation}
When we set all exposure gains $v_e = 1$, all buying gains $v_b = 0$, and each provider's expected gain $y$ to its averaged relevance to users (Eq.~(\ref{eq:grouprel})), our gain-based fairness metric transforms into the aforementioned exposure fairness framework directly. 
However, because existing exposure fairness frameworks assume that all providers have the same needs for the ranking platform (i.e., receiving more exposure in rankings), they essentially ignore the differences between providers and only focus on the equality perspective of fairness.
In contrast, our gain-based provider fairness can better analyze ranking systems from both equity and equality.
To optimize such an equity-oriented gain-based fairness metric, we propose a ranking algorithm: \textbf{EquityRank}.

%% file: tabs/notation.tex
\begin{table}[t]
    \centering
    \caption{A summary of the notations.}
    \begin{tabular}{p{0.065\textwidth} | p{0.35\textwidth}}\hline
    $n$, $\mathcal{I}$, $\tau$, $\mathcal{U}$, $u$ & $\mathcal{I}$ is the global item set contains $n = |\mathcal{I}|$ items, and $\tau \in \mathcal{I}$ indicates a specific item. $\mathcal{U}$ is the user set and $u \in \mathcal{U}$ represents a user.\\\hline
    $\pi, K, k_c, T$ & $\pi$ is a rank list, and its size is $K$. $k_c$ is the cutoff prefix for evaluation and $k_c \leq K$. $T$ denotes a specific time step in the online ranking process.\\\hline
    $m, \mathcal{G}, g$ & $\mathcal{G}$ is the set of all providers $g$, with a size of $m$. \\\hline
    $n_g, \mathcal{I}_{g}$ & $\mathcal{I}_{g}$ is the item set corresponding to provider $g$, with a size of $n_g$. \\\hline
    $v_e^g, v_b^g, y^g$ & $v_e^g$ and $v_b^g$ represent the gain obtained by provider $g$ when a user examine or buy an item from $g_i$, and $y^g$ is the expected gain (ratio) for $g$. \\\hline
    $\mathbf{e}$, $\mathbf{r}$, $\mathbf{b}$ & Binary random variables that indicate if an item is examined by the user ($\mathbf{e}$), relevant to the user ($\mathbf{r}$), and bought by the user ($\mathbf{b}$).\\\hline
    $p_k, r, E, G$ & $p_k$ is the probability of users examining items on position $k$. $r$ is the probability of relevance of an item. $E$ and $G$ denote the exposure and gain of a provider, respectively.  \\\hline
    \end{tabular}
    \label{tab:notations}
\end{table}

%% file: secs/4method.tex
\section{EquityRank}
\label{sec:method}
In this section, we first delve into the ranking objective and then introduce our algorithm accordingly.

\subsection{Ranking Objective}
The ultimate goal of a fair ranking algorithm is to jointly optimize user-side utility (effectiveness) and provider-side fairness, achieving a good balance between the two. 
Specifically, we define the optimization objective of a ranking algorithm with a hyperparameter $\alpha$ to balance the two factors:
\begin{equation}
\max_{\pi} \ \mathrm{Obj.}(T) = \mathrm{eff.}(T) + \alpha \cdot \mathrm{fair.}(T).
\end{equation}

The objective function is non-convex (due to the quadratic terms in $\mathrm{fair.}$) and combinatorial (due to discrete ranking slots), its global optimal solution requires enumerating all permutations, which is infeasible for large-scale systems. 
To this end, we have to propose an efficient approximate method.
As discussed by~\citet{yang2023marginal}, the ranking result at time step $T$ does not influence the previous $\mathrm{Obj.}(T-1)$, so the optimization of the above ranking objective at $T$ is equivalent to maximizing the marginal objective. 
Therefore, we can use the marginal exposure $\Delta E$ to make a first-order approximation of the marginal objective:
\begin{equation}\begin{aligned}
\label{eq:inobj}
\max_{\pi} \  \mathrm{Obj.}(T) & \equiv \max_{\pi} \quad \mathrm{Obj.}(T) - \mathrm{Obj.}(T-1) \\ 
& = \max_{\pi} \quad \Delta \mathrm{Obj.}(T) \\
& = \max_{\pi} \quad \Delta \mathrm{eff.}(T) + \alpha \Delta \mathrm{fair.}(T)  \\
& \approx \max_{\pi} \underbrace{\sum_{\tau \in \mathcal{I}} \frac{\partial \ \mathrm{eff.}(T)}{\partial \ E(\tau)} \Delta E(\tau)}_{\text{Effectiveness}} + \alpha \underbrace{\sum_{\tau \in \mathcal{I}} \frac{\partial \ \mathrm{fair.}(T)}{\partial \ E(\tau)} \Delta E(\tau)}_{\text{Fairness}}.
\end{aligned}\end{equation}

Here we can calculate the derivatives for the two terms in Eq.~(\ref{eq:inobj}) separately. First, for the effectiveness term (which is measured by aNDCG or cNDCG as discussed in~\S\ref{subsec:user_effect}), we disregard the discounted factor $\gamma$ (by setting it to 1) and the denominator employed for normalization (i.e., the constant IDCG value). This allows us to reformulate it with the definition of exposure function (Eq.~(\ref{eq:exposure})) as:
\begin{equation}\begin{aligned}
\mathrm{eff.}(T) & = \sum_{t=1}^T\sum_{k=1}^{K} r(\pi_t[k]) \cdot p_k \\ 
& = \sum_{\tau \in \mathcal{I}} \sum_{t=1}^T\sum_{k=1}^{K} \mathbb{I}[\pi_t[k] == \tau] \cdot r(\tau) \cdot p_k \\ 
& = \sum_{\tau \in \mathcal{I}} r(\tau) \sum_{t=1}^T\sum_{k=1}^{K} \mathbb{I}[\pi_t[k] == \tau] \cdot p_k \\
& = \sum_{\tau \in \mathcal{I}} r(\tau) E(\tau).
\end{aligned}\end{equation}
Therefore, the derivative of the effectiveness function is:
\begin{equation}
\frac{\partial \ \mathrm{eff.}(T)}{\partial \ E(\tau)} = r(\tau).
\end{equation}

On the other hand, for the fairness term, directly calculating the derivative is challenging. Thus, we use the chain rule by first calculating the derivative of fairness $\mathrm{fair.}$ with respect to the gain of each provider $G(g)$ (Eq.~(\ref{eq:gain})), and then the derivative of gain with respect to the item exposure:
\begin{equation}
\frac{\partial \ \mathrm{fair.}(T)}{\partial \ E(\tau)}  = \sum_{g \in \mathcal{G}} \frac{\partial \ \mathrm{fair.}(T)}{\partial \ G(g)} \cdot \frac{\partial \ G(g)}{\partial \ E(\tau)}.
\end{equation}
The first term can be easily derived based on the definition of fairness (Eq.~(\ref{eq:fair}), Eq.~(\ref{eq:unfair})). We denote this result as $B(g_i)$:
\begin{equation}\label{eq:fair_grad}
\frac{\partial \ \mathrm{fair.}(T)}{\partial \ G(g)} = B(g) = \frac{4}{m(m-1)}\left(y^{g}\sum_{g'}G(g')y^{g'} - G(g)\sum_{g'}(y^{g'})^2\right). 
\end{equation}
Before calculating the derivative of $G$, let's first rewrite it (based on Eq.~(\ref{eq:gain_pertime}) and neglect the denominator $1/T$ in Eq.~(\ref{eq:gain})):

\begin{equation}\begin{aligned}
G(g) & = \sum_{t=1}^T \sum_{\tau \in \mathcal{I}_{g} \cap \pi_t } p_{\rm{rank}(\tau | \pi_t)} (v_e^{g} + r(\tau) v_b^{g}) \\ 
& =\sum_{t=1}^T \sum_{\tau \in \mathcal{I}} \sum_{k=1}^{K} \mathbb{I}[\pi_t[k] == \tau \land \tau \in \mathcal{I}_{g}] \cdot p_k \cdot \left(v_e^{g} + r(\tau)v_b^{g} \right) \\ 
& = \sum_{\tau \in \mathcal{I}}\mathbb{I}[\tau \in \mathcal{I}_{g}] \cdot \left(v_e^{g} + r(\tau)v_b^{g}\right) \sum_{t=1}^T\sum_{k=1}^{K} \mathbb{I}[\pi_t[k] == \tau] p_k \\ 
& = \sum_{\tau \in \mathcal{I}}\mathbb{I}[\tau \in \mathcal{I}_{g}] \cdot \left(v_e^{g} + r(\tau)v_b^{g}\right) \cdot E(\tau).
\end{aligned}\end{equation}
Thus the derivative of the gain with respect to an item's exposure can be expressed as:
\begin{equation}\begin{aligned}
\label{eq:gain_grad}
\frac{\partial \ G(g)}{\partial \ E(\tau)} = \begin{cases}v_e^{g} + r(\tau) \cdot v_b^{g}, & \text { if } \tau \in \mathcal{I}_{g},\\ 0, & \text { otherwise}. \end{cases}
\end{aligned}\end{equation}

Combining Eq.~(\ref{eq:inobj}) through Eq.~(\ref{eq:gain_grad}), we can derive the ultimate ranking objective at each time step:
\begin{equation}
\label{eq:final_obj}
\max \sum_{\tau \in \mathcal{I}} \underbrace{\left[r(\tau) + \alpha \cdot B({\rm{Group}(\tau)})\left(v_e^{\rm{Group}(\tau)} + r(\tau)v_b^{\rm{Group}(\tau)}\right)\right]}_{\mathrm{grad}(\tau)} \cdot \Delta E(\tau),
\end{equation}
where $\rm{Group}(\cdot)$ indicates which provider or group an item belongs to, while we use $\mathrm{grad}(\cdot)$ to represent the overall gradient of an item.

\begin{figure}[t]
    \centering
    \includegraphics[width=0.46\textwidth]{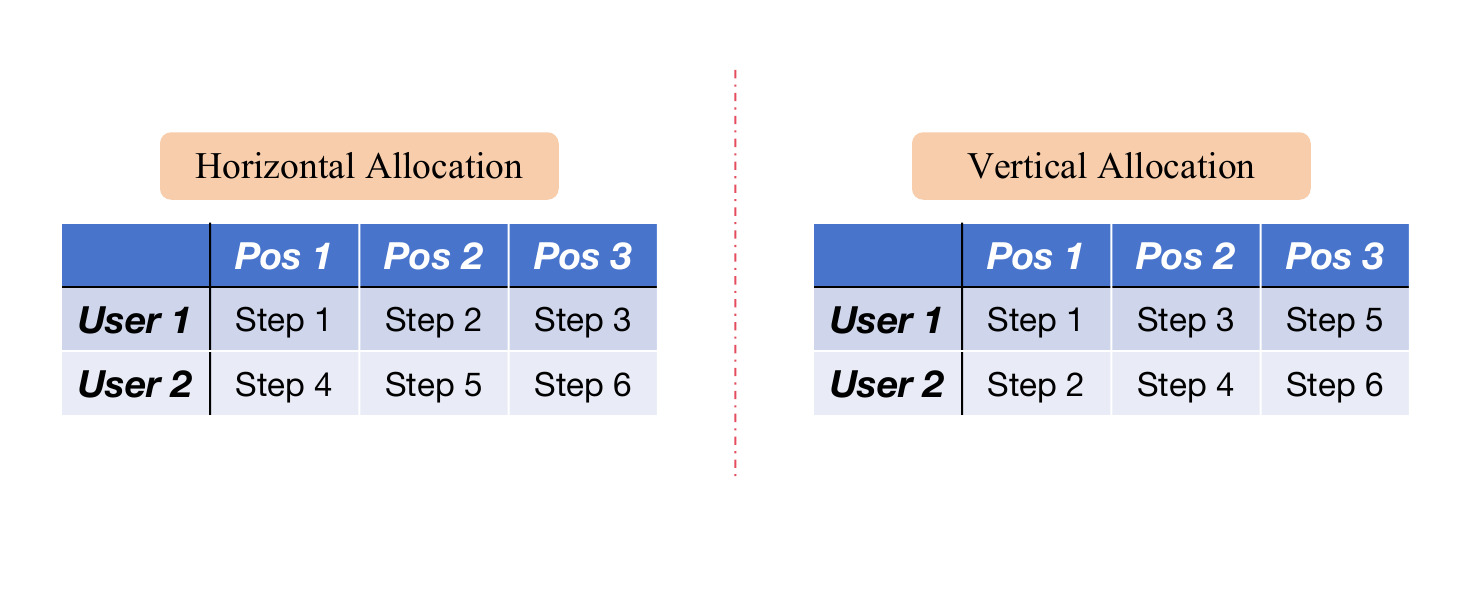}
    \vspace{-6mm}
    \caption{A schematic diagram of horizontal and vertical allocation~\cite{yang2023vertical} with $2$ users and rank list size $K = 3$.}
    \label{fig:ver}
\end{figure}

\subsection{Ranking Strategy \& Efficiency}
Now we can formulate the ranking strategy after determining the objective function. 
Intuitively, we should prioritize items with larger gradients $\mathrm{grad}$, placing them in higher positions. 
Therefore, in the online setting where each user comes to the ranking platform on the fly, we sort items based on their $\mathrm{grad}$ and show them to each user directly.
In the offline setting where relevance labels and users are already given and multiple rank lists have to be generated at once, we use the vertical allocation method (see Figure~\ref{fig:ver}) instead of horizontal allocation to achieve better performance in higher positions~\cite{yang2023vertical}. 
Specifically, we select the item with the largest current gradient at each step. 

Then we analyze the time complexity of EquityRank, particularly focusing on the online setting where computational efficiency is critical for on-the-fly service.
In each iteration, it optimizes a gain-based objective consisting of an effectiveness term and a fairness-correction term (Eq.~(\ref{eq:inobj})).
For effectiveness, EquityRank directly leverages pre-computed relevance judgements without additional calculation.
For fairness, it computes the group-wise fairness gradients $B(g)$ for each group $g$ (Eq.~(\ref{eq:fair_grad})) involving pairwise comparisons among all $m$ provider groups and yields a time complexity of $\mathcal{O}(m^2)$. 
Then, for each candidate, it assembles the calculated fairness gradient $B$ with the relevance judgment to assign a final score (i.e., the overall gradient $\mathrm{grad}(\tau)$ in Eq.~(\ref{eq:final_obj})) to each of the $n$ items. 
Since we can store the group information in a hash table, this step has a total complexity of $\mathcal{O}(n)$.
Finally, EquityRank sorts all items based on their overall scores $\mathrm{grad}(\tau)$, which incurs a time complexity of $\mathcal{O}(n \log n)$ using an efficient comparison-based algorithm like quicksort.
Therefore, the overall time complexity for one online step of is determined by the dominant factors: $\mathcal{O}(n \log n + m^2)$, which is the same as the previous baselines like FairCo~\cite{morik2020controlling} (detailed in \S\ref{subsec:baseline}). 
On the other hand, in practice, the number of provider groups $m$ is typically much smaller than the number of items $n$, so the $\mathcal{O}(n \log n)$ term dominates the overall complexity.

%% file: secs/5experi.tex
\section{Experimental Setup}\label{sec:experiment}
\subsection{Datasets and Data Process}\label{subsubsec:data}
In our experiment, we utilize four large-scale and publicly available recommender datasets, including:
\begin{itemize}[leftmargin=*]
    \item \textbf{Musical Instruments~(MI)} and \textbf{Video Games~(VG)}\footnote{\url{https://cseweb.ucsd.edu/~jmcauley/datasets/amazon_v2/}} are two subsets of the Amazon Product Reviews dataset~\cite{ni2019justifying} that includes reviews and product metadata from Amazon. 
    Following \cite{xu2023p,xu2024taxation}, we only consider the clicked data simulated as the 4-5 star rating samples. 
    Brands are considered as providers.
    \item \textbf{Alaska~(AL)}\footnote{\url{https://datarepo.eng.ucsd.edu/mcauley_group/gdrive/googlelocal/}} is a subset of the Google Local dataset~\cite{li2022uctopic} about business reviews. 
    Also, only clicked data simulated as 4-5 star rating samples are employed. 
    The first category of the business is considered as the provider.
    \item \textbf{RateBeer~(RB)}\footnote{\url{https://datarepo.eng.ucsd.edu/mcauley_group/data/beer/ratebeer.json.gz}} contains beer reviews with multiple rated dimensions~\cite{he2016vbpr}. 
    Similarly, 16-20 overall rating samples are regarded as clicked samples and different brewers are considered as providers.
\end{itemize}
Following \cite{xu2023p,xu2024taxation}, users/items whose interaction counts are fewer than 10 and providers that are associated with fewer than 20 items are removed in advance. 
All dataset features are outlined in Table~\ref{tab:Dataset_statistic}.

On each dataset, we first train a NeuMF model~\cite{he2017neural} via  RecBole\footnote{\url{https://github.com/RUCAIBox/RecBole}}  with the default parameter setting to generate the relevance score for each user-item pair. 
Since we cannot access the real provider information from the e-commerce platforms for each dataset, we simulate providers' income and expected gain~(i.e., $v_e, v_b, y$) through random sampling.
Based on the common observation that purchase income should significantly exceed exposure income~\cite{yang2020search}, meanwhile there are also significant differences between providers (such as large enterprises and small-scale vendors), we conduct the sampling as follows: $v_e\sim \mathcal{N}(10, 2.5)$~(i.e., normal distribution with mean as 10 and standard deviation as 2.5), $v_b\sim \mathcal{N}(100, 25)$, and $y\sim \mathcal{N}(50, 25)$.
We refer to this simulation scenario as Common.
To conduct a more comprehensive comparison and further validate the effectiveness of our method, we consider two special scenarios that may occur in practice.
(1) There are scenarios where merchants focus more on product promotion and exposure to attract users' attention and increase their brand awareness as well as customer loyalty, while not aiming to sell large quantities of products within a short period.
In this case, we assume that the gain from item exposure would be significantly larger than common, and sample both $v_e$ and $v_b$ from $\mathcal{N}(100, 25)$.
We refer to this scenario as \textbf{Exp1st}.
(2) There are time periods, such as during shopping festivals like Amazon Black Friday and Taobao 11.11 Sale, when users tend to purchase large quantities of goods in bulk.
In this scenario, merchants aim to sell as many products as possible and as quickly as possible, capitalizing on the favorable impressions created during prior promotional activities.
Here, we sample $v_e\sim \mathcal{N}(10, 2.5), v_b\sim \mathcal{N}(1000, 250)$ to simulate this case and refer to it as \textbf{Sale1st}.

\begin{table}[t]
    \centering
    \caption{A summary of the dataset features.}
    \resizebox{0.9\columnwidth}{!}{
    \begin{tabular}{c|ccc}
    \toprule
    Dataset / Property & Users & Items & Providers\\
    \midrule
    Amazon-Musical Instruments~(MI) & 4,801 & 5,418 & 105\\
    Amazon-Video Games~(VG) & 8,966 & 12,351 & 115\\
    Google Local-Alaska~(AL) & 16,651 & 3,511 & 72\\
    RateBeer~(RB) & 12,293 & 22,349 & 533 \\
    \bottomrule
    \end{tabular}
    }
    \label{tab:Dataset_statistic}
\end{table}

\subsection{Baselines}\label{subsec:baseline}
To demonstrate the effectiveness of our proposed EquityRank algorithm, we compare it to two naive but intuitive methods and three state-of-the-art group-wise fair ranking algorithms:
\begin{itemize}[leftmargin=*]
    \item \textbf{TopK} Rank items simply according to the relevance.
    \item \textbf{PoorK} First select a group based on the imbalance ratio $\frac{G(g)}{y^g}$, then choose the most relevant item within that group.
    \item \textbf{FairCo}~\cite{morik2020controlling} Use a proportional controller to balance relevance and exposure fairness. We also implement it under our equity-based fairness metric as \textbf{FairCo*}.
    \item \textbf{MMF}~\cite{yang2021maximizing} A method based on random sampling focuses on exposure fairness at the top positions. We also implement it under our equity-based fairness metric as \textbf{MMF*}. When the balancing parameter $\alpha=1$, \textbf{MMF*} degrades to \textbf{PoorK}. 
    \item \textbf{TFROM}~\cite{wu2021tfrom} A ranking algorithm considers both provider-side exposure fairness and user-side fairness. We also implement it under our equity-based fairness metric as \textbf{TFROM*}.
    \item \textbf{EquityRank} and \textbf{EquityRank\_v} Our methods. Sort items according to the gradient $\mathrm{grad}(\tau)$ in Eq.~(\ref{eq:final_obj}). EquityRank\_v denotes using vertical allocation in Figure~\ref{fig:ver} for offline settings.
\end{itemize}
Among these models, FairCo, MMF, and our EquityRank employ a hyperparameter $\alpha$ to balance relevance and fairness. 
Rather than tuning it, we thoroughly compare across the parameter space by running each model with various $\alpha$. 
This allows us to compare these models in distinct scenarios where the ranking system has a varying emphasis on the two factors.
Note that models like FairRec~\cite{patro2020fairrec} and P-MMF~\cite{xu2023p} address the problem of fair allocation rather than fair ranking. 
That is, they treat all slots in the list as equivalent and do not account for position bias, making them incompatible with our framework and thus not included in the comparison.

\input{tabs/offline_expo}

\subsection{Ranking Service Simulation}
As outlined in ~\S\ref{subsec:rank_service}, the workflow involves randomly selecting a simulated user at each time step. 
Subsequently, the ranking algorithm generates a ranked list (with size $K = 5$) for the user, presents it to the user for feedback, and finally gains from the interaction.
Our experiments can be generally categorized into two settings: offline and online. 
In the offline setting, true relevance labels are already given, the number of users is fixed, and all ranking results are simultaneously shown to users at once based on expected gains (Eq.~(\ref{eq:gain_pertime})). 
In contrast, the online setting resembles search engines or recommendation systems, requiring immediate rank results for each user. 
In the online setting, we simulate the workflow for $2.5\times10^5$ steps, serving one user per step. 
Here, exposure gain can be computed directly using the total exposure $E$ harvested by a provider. 
Regarding purchase behavior, we simulate whether the user buys the current item by Bernoulli sampling using $P(\mathbf{b} = 1)$, and the provider can only receive the purchase gain if the user opts to buy the item (i.e., $\mathbf{b} = 1$). 
To simulate user interactions and provider gains mentioned above, we also need to model exposure and relevance.
For exposure calculation, following the approach in~\cite{oosterhuis2021unifying,yang2023marginal}, the user's examination probability is calculated as $p_k = \frac{1}{\rm{log}_2(k) + 1}$. 
For relevance modeling, we adopt different methods for offline and online ranking settings. 
In the offline setting, we assume that the relevance model is perfectly trained beforehand and directly use the true relevance labels generated in \S\ref{subsubsec:data} as the probability of relevance $r(\tau)$ estimated by the relevance model. 
However, in the online setting, the relevance model has to learn the relevance based on the user's feedback during the ranking process. 
We refer to previous works~\cite{oosterhuis2021unifying,yang2023marginal} and use an unbiased estimator: $\hat{r}(\tau) = \frac{cumB(\tau)}{E(\tau)}$ where $cumB$ represents a specific user's cumulative purchases of a particular item. 
Thus, we replace $r(\tau)$ to $\hat{r}(\tau)$ in the online ranking objective of Eq.~(\ref{eq:final_obj}). 
It is important to note that in the online setting, starting the system from scratch with a vast number of items could be difficult to converge. 
Thus, we train a weaker NeuMF model (without the MLP module) to filter the top 20 relevant items for each user (akin to the coarse-grained ranking phase in a practical IR system). 
We only conduct fair re-ranking on these top items which actually cover an average of 94\% of all original items. 

\begin{figure}[t]
    \centering
    \includegraphics[width=0.4\textwidth]{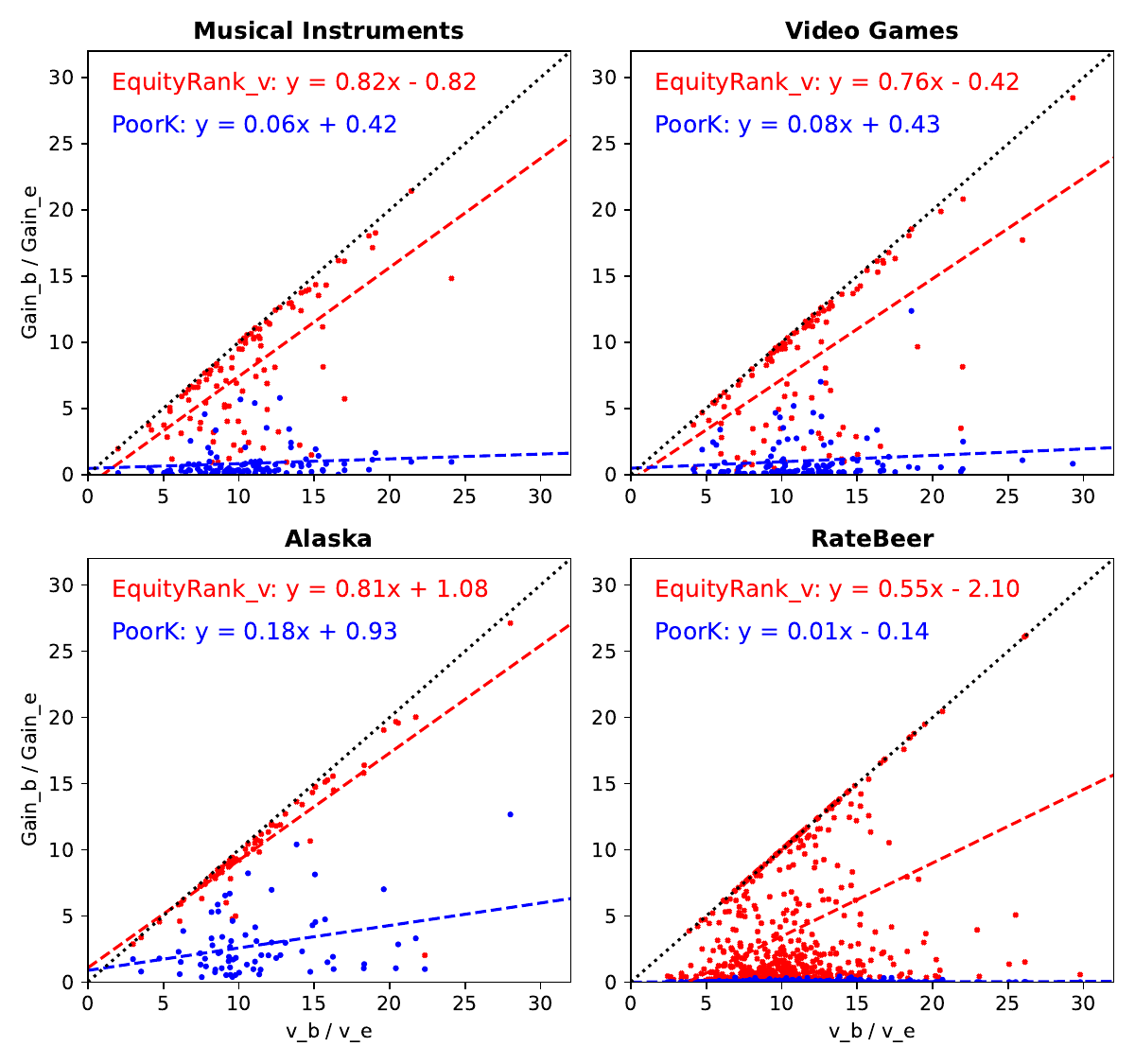}
    \caption{A comparison of the fairness ranking results between EquityRank\_v and PoorK. We first calculate the ratio of the two types of gain weights for each provider in each dataset, $\frac{v_b}{v_e}$. Then, we compute the ratio of the two types of gains assigned to each provider by the two models, $\frac{Gain_b}{Gain_e}$. Each point represents a provider's own $\frac{v_b}{v_e}$ (the x-axis) and its allocated $\frac{Gain_b}{Gain_e}$ under a specific algorithm (the y-axis).}
    \label{fig:ratio}
\end{figure}

\begin{table}[t]
    \centering
    \caption{The mean squared difference~(MSD$\downarrow$) and Pearson correlation coefficienton~($\rho$$\uparrow$) between the gain allocation of the providers~($\frac{Gain_b}{Gain_e}$) by each model and the expected ratio of these providers~($\frac{v_b}{v_e}$). "*" and "$\dagger$" denote significantly worse performance than our EquityRank and EquityRank\_v with $p < 0.05$ level, respectively.}
    \resizebox{0.95\columnwidth}{!}{
    \begin{tabular}{c||cc|cc|cc|cc}
    \toprule
    \multirow{2}{*}{Models} & \multicolumn{2}{c|}{MI} & \multicolumn{2}{c|}{VG} & \multicolumn{2}{c|}{AL} & \multicolumn{2}{c}{RB} \\ \cline{2-9} 
     & MSD$\downarrow$ & $\rho$$\uparrow$ & MSD$\downarrow$ & $\rho$$\uparrow$ & MSD$\downarrow$ & $\rho$$\uparrow$ & MSD$\downarrow$ & $\rho$$\uparrow$ \\ 
     \hline
    FairCo* & $117.3^{*\dagger}$ & $0.06^{*\dagger}$ & $136.5^{*\dagger}$ & $0.08^{*\dagger}$ & $103.8^{*\dagger}$ & $0.25^{*\dagger}$ & $134.6^{*\dagger}$ & $0.03^{*\dagger}$ \\
    PoorK/MMF* & $110.5^{*\dagger}$ & $0.12^{*\dagger}$ & $131.2^{*\dagger}$ & $0.13^{*\dagger}$ & $92.5^{*\dagger}$ & $0.32^{*\dagger}$ & $137.3^{*\dagger}$ & $0.12^{*\dagger}$ \\
    TFROM* & $123.2^{*\dagger}$ & $0.05^{*\dagger}$ & $151.2^{*\dagger}$ & $-0.12^{*\dagger}$ & $142.9^{*\dagger}$ & $-0.01^{*\dagger}$ & $135.7^{*\dagger}$ & $0.02^{*\dagger}$ \\
    EquityRank & $29.7^\dagger$ & $0.58^\dagger$ & $37.0^\dagger$ & $0.59^\dagger$ & $12.7^\dagger$ & $0.79^\dagger$ & $97.4^\dagger$ & $0.31^\dagger$ \\
    EquityRank\_v & \textbf{18.1} & \textbf{0.69} & \textbf{28.0} & \textbf{0.63} & \textbf{7.3} & \textbf{0.85} & \textbf{72.6} & \textbf{0.51} \\
    \bottomrule
    \end{tabular}
    }
    \label{tab:ratio}
\end{table}

\subsection{Evaluation}
We use the aNDCG and cNDCG (Eq.~(\ref{eq:acndcg}), with $\gamma = 0.995, k_c=K$) with the relevance label generated in~\S\ref{subsubsec:data} to assess effectiveness. 
The definition in Eq.~(\ref{eq:unfair}) is employed to measure provider-side unfairness.
We run each experiment five times and report the average performance on a single \textit{Intel Xeon Gold 5218} CPU.

%% file: tabs/offline_expo.tex
\begin{table*}[t]
    \centering
    \caption{The minimum unfairness value~($\downarrow$) each model can achieve in the offline setting. Standard deviation is in the parentheses. We emphasize the best result in each dataset. We also report the average processing time for each algorithm over 100 samples. For FairCo, MMF, and TFROM, the versions without "*" correspond to the original exposure-fairness settings, whereas those marked with "*" denote the variants modified according to our proposed equity-oriented fairness formulation.}
    \resizebox{0.8\textwidth}{!}{
    \begin{tabular}{c||cc|cc|cc|cc}
    \toprule
    \multirow{3}{*}{Models} & \multicolumn{8}{c}{Datasets} \\
    \cline{2-9} & \multicolumn{2}{c|}{MI} & \multicolumn{2}{c|}{VG} & \multicolumn{2}{c|}{AL} & \multicolumn{2}{c}{RB} \\
    \cline{2-9} & unfair. & time (sec) & unfair. & time (sec) & unfair. & time (sec) & unfair. & time (sec) \\
    \midrule
    TopK & $1.38\times10^5_{(5.2\times10^3)}$ & 0.159 & 
    $2.10\times10^5_{(7.5\times10^3)}$ & 0.318 & $1.85\times10^5_{(6.8\times10^3)}$ & 0.109 & $1.92\times10^3_{(94)}$ & 0.493 \\
    PoorK & $0.126_{(7.1\times10^{-3})}$ & 0.146 & $0.043_{(3.7\times10^{-3})}$ & 0.268 &  $0.019_{(2.4\times10^{-3})}$ & 0.101 & $\mathbf{0.002}_{(2.1\times10^{-4})}$ & 0.460 \\
    \midrule
    FairCo & $4.02\times10^3_{(2.1\times10^2)}$ & 0.207 & $3.17\times10^3_{(1.7\times10^2)}$ & 0.418 & $2.42\times10^4_{(9.6\times10^2)}$ & 0.146 & $9.96_{(0.63)}$ & 0.670 \\
    MMF & $1.12\times10^4_{(6.2\times10^2)}$ & 1.903 & $7.28\times10^3_{(4.1\times10^2)}$ & 3.725 & $5.86\times10^4_{(3.3\times10^3)}$ & 1.708 & $10.8_{(0.59)}$ & 8.614 \\
    TFROM & $1.38\times10^5_{(5.2\times10^3)}$ & 0.260 & $2.10\times10^5_{(7.5\times10^3)}$ & 0.596 & $1.85\times10^5_{(6.8\times10^3)}$ & 0.233 & $1.92\times10^3_{(94)}$ & 1.006 \\
    \midrule
    FairCo* & $0.285_{(1.8\times10^{-2})}$ & 0.213 & $0.093_{(7.6\times10^{-3})}$ & 0.412 & $0.038_{(4.0\times10^{-3})}$ & 0.139 & $0.022_{(2.7\times10^{-3})}$ & 0.688 \\
    MMF* & $0.126_{(7.1\times10^{-3})}$ & 1.922 & $0.043_{(3.7\times10^{-3})}$ & 4.074 & $0.019_{(2.4\times10^{-3})}$ & 1.747 & $\mathbf{0.002}_{(2.1\times10^{-4})}$ & 9.203\\
    TFROM* & $0.106_{(8.3\times10^{-3})}$ & 0.283 & $0.056_{(5.0\times10^{-3})}$ & 0.563 & $0.013_{(1.6\times10^{-3})}$ & 0.254 & $0.046_{(4.8\times10^{-3})}$ & 1.047 \\
    \midrule
    EquityRank (ours) & $0.499_{(2.3\times10^{-2})}$ & 0.226 & $0.151_{(9.5\times10^{-3})}$ & 0.395 & $0.037_{(3.8\times10^{-3})}$ & 0.131 & $0.103_{(6.2\times10^{-3})}$ & 0.596 \\
    EquityRank\_v (ours) & $\mathbf{0.101}_{(5.6\times10^{-3})}$ & 0.421 & $\mathbf{0.037}_{(3.0\times10^{-3})}$ & 0.722 & $\mathbf{0.012}_{(1.6\times10^{-3})}$ & 0.295 & $0.024_{(2.6\times10^{-3})}$ & 1.165 \\
    \bottomrule
    \end{tabular}
    }
    \label{tab:offline_expo}
\end{table*}

%% file: secs/6result.tex
\begin{figure*}[t]
    \centering
    \includegraphics[width=0.85\textwidth]{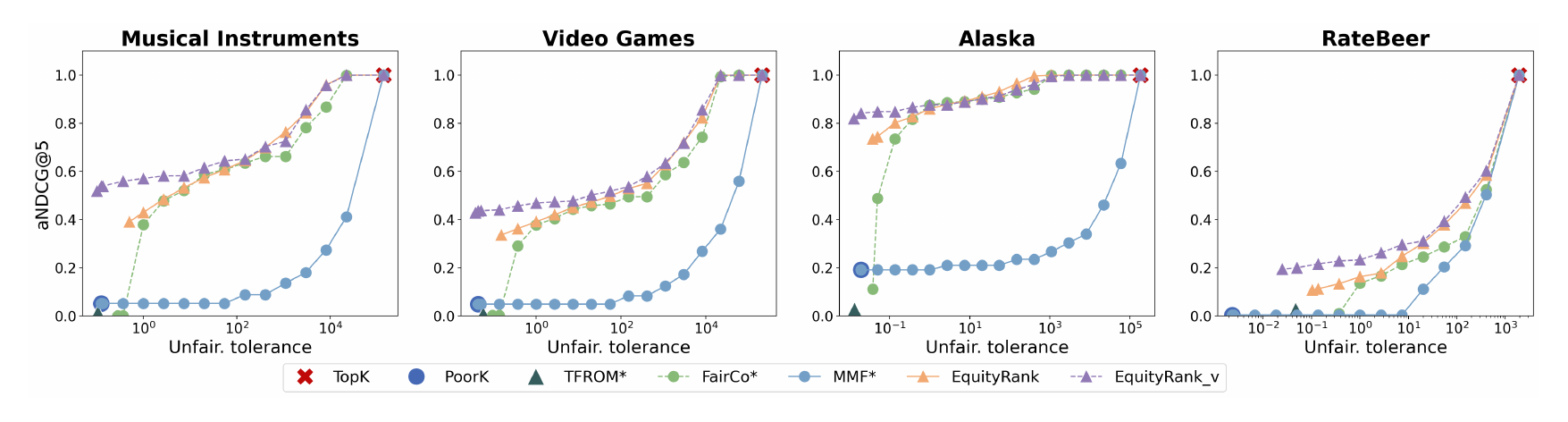}
    \caption{The unfair.-aNDCG balance curve for each model across different datasets under the offline setting. For models with a hyperparameter $\alpha$, it indicates the highest NDCG an algorithm can achieve when the unfairness value does not exceed a certain threshold. The upper and lefter ($\nwarrow$) a point, the better its corresponding model's performance.}
    \label{fig:offline_balance}
\end{figure*}

\begin{figure}[t]
    \centering
    \includegraphics[width=0.4\textwidth]{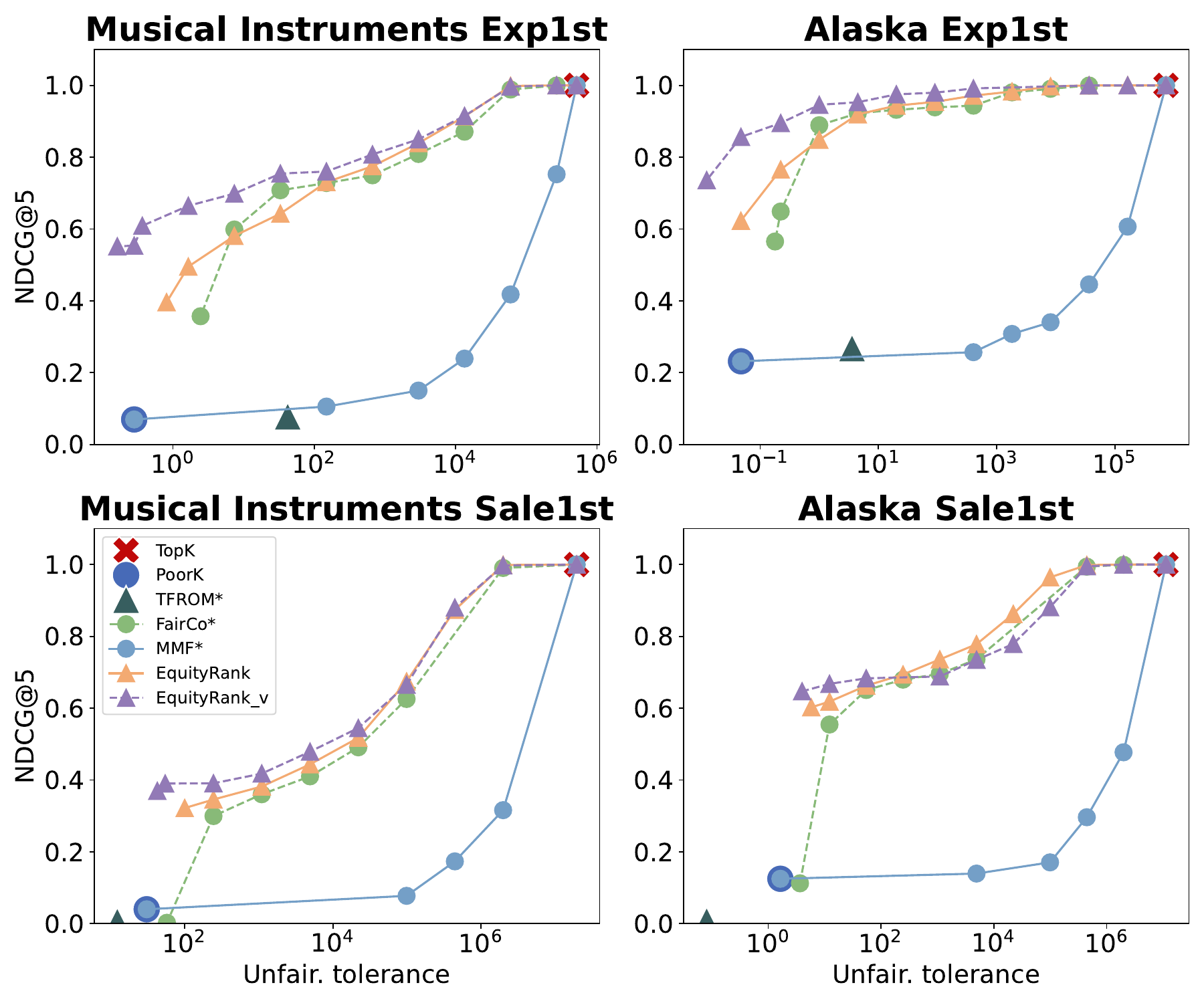}
    \caption{The unfair.-aNDCG curves in the two special scenarios under the offline setting. "Exp1st" and "Sale1st" represent providers preferring exposure and sales, respectively.}
    \label{fig:offline_extreme}
\end{figure}

\begin{figure*}[t]
    \centering
    \includegraphics[width=0.85\textwidth]{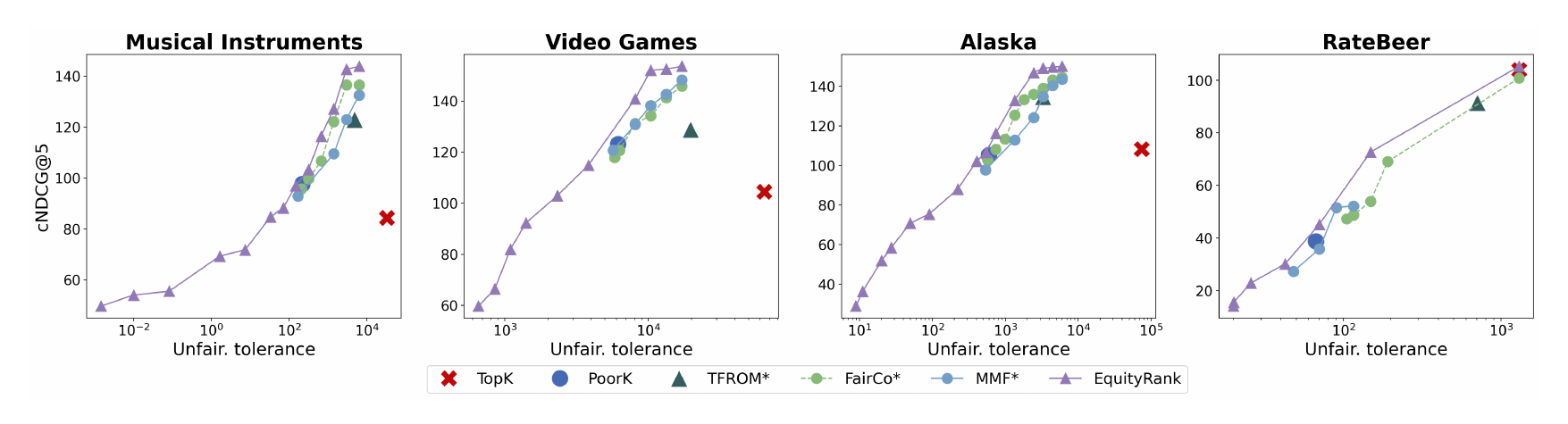}
    \caption{The unfair.-cNDCG balance curve for each model across different datasets under the online setting.}
    \label{fig:online}
\end{figure*}

\begin{figure}[t]
    \centering
    \includegraphics[width=0.4\textwidth]{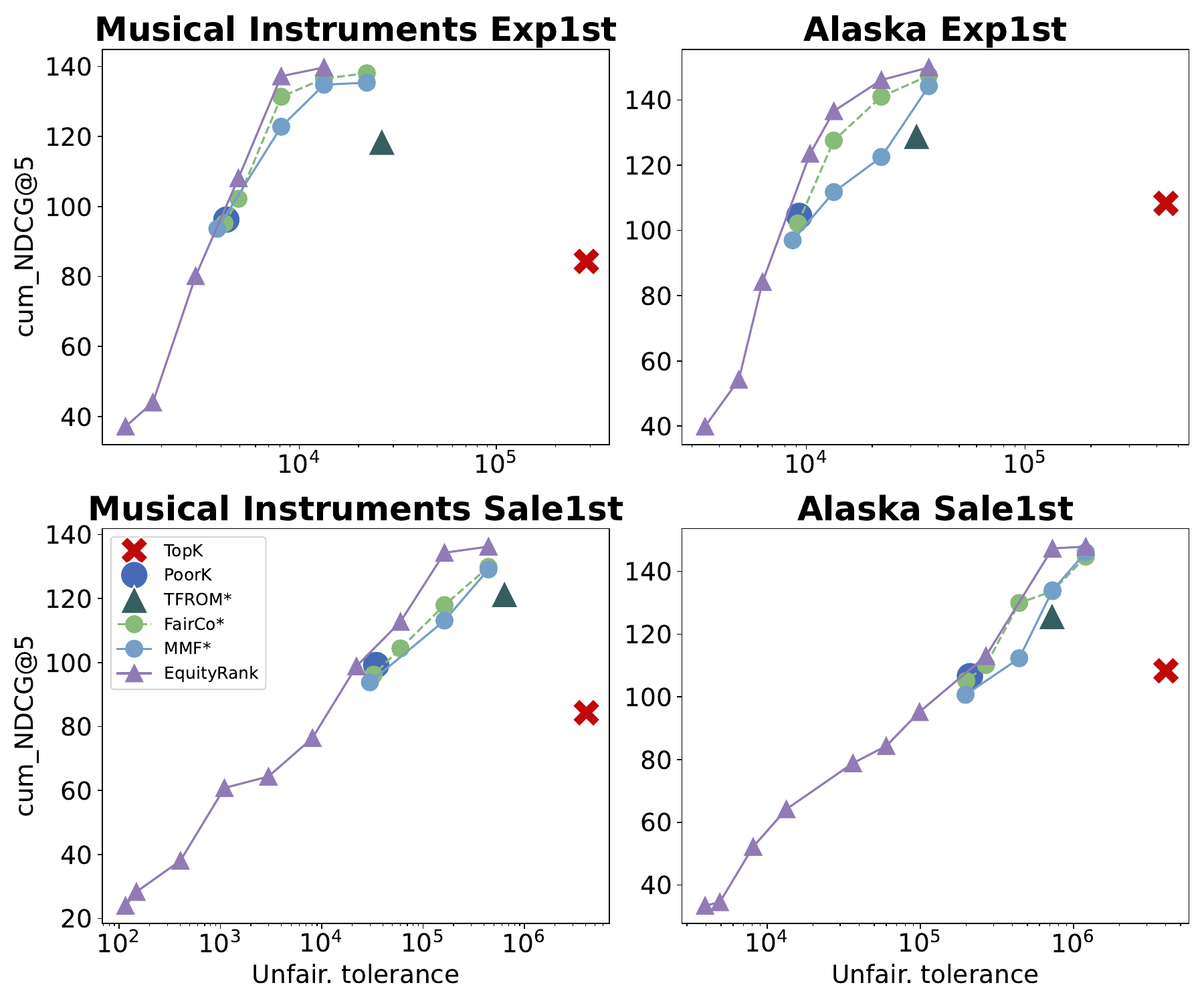}
    \caption{The unfair.-cNDCG curves of the two special scenarios under the online setting. "Exp1st" and "Sale1st" represent providers preferring exposure and sales, respectively.}
    \label{fig:online_extreme}
\end{figure}

\section{Results and Analysis}
\subsection{Offline Performance on Equity-oriented Fairness}
Table~\ref{tab:offline_expo} compares the minimum unfairness values that different algorithms can achieve in the offline setting. 
In this comparison, equality-oriented algorithms, FairCo, MMF, and TFROM, temporarily use the original exposure-based fairness definition for optimization. 
The algorithm without fairness consideration, TopK, as well as these exposure-based methods, fail to fully comprehend the diverse needs of different providers, and thus cannot ensure fairness in terms of provider benefits, ultimately failing to achieve equity. 
In contrast, equity-based fair methods can effectively optimize for equity-oriented fairness objectives. 
Especially, our EquityRank\_v significantly outperforms other baselines within an acceptable time cost on most datasets except RB.

To further demonstrate that EquityRank can better meet the diverse needs of different providers and achieve equity, we compare the allocation of purchase gains and exposure gains to different providers by different algorithms.
For each dataset, we first compute the ratio between two types of gain weights $\frac{v_b}{v_e}$ for each provider, then its gain ratio allocated by one algorithm $\frac{Gain_b}{Gain_e}$. 
Subsequently, we compute the mean squared difference~(MSD) and the Pearson correlation coefficient~($\rho$) between the ratios allocated by each equity-oriented approach and the original data ratios (i.e., $\frac{1}{m}\sum_{g \in \mathcal{G}}(\frac{Gain_b^g}{Gain_e^g}-\frac{v_b^g}{v_e^g})^2$ and $\rho(\frac{Gain_b}{Gain_e}, \frac{v_b}{v_e})$). 
A lower MSD and a higher $\rho$ value indicate that the different types of profits obtained by providers better meet their tailored demand.
Table~\ref{tab:ratio} illustrates that our EquityRank and especially EquityRank\_v can generate ranking results that better align with provider needs by accounting for each provider's specific preferences for different factors.
In other words, they tend to assign a greater proportion of purchase benefits to providers who prioritize purchase incomes and vice versa, thereby indicating that our algorithm better fulfills their preferences on different factors and diverse needs. 
We also conduct a case study between EquityRank\_v and PoorK in Figure~\ref{fig:ratio}, with $\frac{v_b}{v_e}$ as the x-axis and $\frac{Gain_b}{Gain_e}$ as the y-axis, representing the expected profit ratio and the actual profit ratio of the provider, respectively. 
The closer a point lies to the line $y = x$, the better the algorithm aligns with the provider's actual preferences.
Figure~\ref{fig:ratio} demonstrates that although the minimum unfairness values of the two algorithms are close, the profits allocated to different providers by EquityRank\_v aligns significantly better with the weight ratios of these providers in the original dataset compared to PoorK (i.e., red points are generally closer to the line $y = x$ compared to blue points). 
Consequently, our algorithm can better cater to providers' varying needs and preferences, thereby enhancing equity-oriented fairness.

\subsection{Effectiveness and Fairness Trade-off} 
Since improving the fairness of ranking results often requires promoting less relevant items to higher positions in the list, it inevitably leads to a certain degree of degradation in relevance.
However, in real-world scenarios, different ranking systems may have various demands on user effectiveness and provider fairness, making the trade-off between both sides a crucial issue.
Therefore, we need to measure the performance in terms of both simultaneously to find a model that better adapts to different conditions. 
To conduct a more comprehensive comparison, in the following experiments, we replace the fairness objective in the exposure-based methods with our equity-oriented measurement, denoted as FairCo*, MMF*, and TFROM*.
To illustrate the performance of different models under varying user effectiveness and provider fairness requirements, we plot the unfairness-aNDCG curve (Figure~\ref{fig:offline_balance}). For models without the balance hyperparameter $\alpha$, their results are represented by single points. 
For models with a trade-off hyperparameter $\alpha$ (FairCo*, MMF*, EquityRank), we test on different hyperparameter settings across the parameter space and plot a curve. 
Each point on the curve represents the highest NDCG value that the model can achieve when the unfairness value does not exceed its corresponding x-axis coordinate. 
Therefore, in the figure, the lefter a point is, the fairer it is; and the higher a point is, the more relevant it is. 

Figure~\ref{fig:offline_balance} illustrates that when the system prioritizes relevance (the right side of each chart), EquityRank, EquityRank\_v, and FairCo consistently outperform other models with minimal differences in their overall performance. 
However, as the fairness requirement intensifies (the left side of each chart, specifically when the unfairness value drops to around $1$ or lower), FairCo’s relevance performance declines sharply, whereas our EquityRank and EquityRank\_v maintain greater stability.
On one hand, our algorithms, especially EquityRank\_v, can approach, reach, or even exceed the fairness level of PoorK and TFROM across all datasets. 
On the other hand, they can maintain high relevance when achieving greater fairness. 
Thus, our methods achieve a superior balance between relevance and fairness, making it more adaptable to the varied needs of platforms and increasing its overall applicability.

Moreover, to assess the model’s applicability across different provider needs, we experiment on two extreme scenarios mentioned in~\S\ref{subsubsec:data}, where providers prioritize either item exposure or high-volume sales.
We run experiments on the ML and AL datasets, and the unfairness-NDCG curves are shown in Figure~\ref{fig:offline_extreme}. 
It indicates that our approaches consistently exhibit strong performance across these scenarios. 
When the ranking system focuses more on relevance, their performance remains among the best, and when fairness is prioritized, they achieve nearly optimal fairness while still maintaining better relevance than other models.

\subsection{Online Results}
Since we have already conducted a detailed analysis and discussion of our EquityRank algorithm in the offline setting, the primary focus of our investigation in the online setting is:
\textit{Can EquityRank still maintain its excellent performance and strong capabilities in the online setting?}
Similar to the offline approaches, we also plot the unfairness-cNDCG curves in the online setting (Figure~\ref{fig:online}) and compare the capability of different models to handle both fairness and relevance. 
Due to the explore-exploit trade-off in the online ranking setting~\cite{wang2018efficient,yue2009interactively,yang2023marginal}, the naive TopK method may not achieve the best relevance. 
In contrast, EquityRank achieves the highest relevance performance across all datasets.
Regarding fairness, it not only attains the optimal fairness scores across all datasets in the online setting, but also tends to achieve higher relevance performance compared to other models when the fairness degree is comparable, with its curve generally positioned in the upper left compared to other models in Figure~\ref{fig:online}, Pareto dominating other ranking strategies. 
Even in the two special scenarios (as shown in Figure~\ref{fig:online_extreme}), EquityRank still demonstrates similarly outstanding performance. Not only can it ensure optimal relevance and fairness performance, but it also more effectively harmonizes user-side effectiveness with provider-side fairness.
As a result, EquityRank remains highly effective in the online setting, successfully balancing user-side relevance and provider-side equity-oriented fairness.

%% file: secs/7conclu.tex
\section{Conclusion and Future Work}
\label{sec:conclusion}
In this paper, we introduce an equity-oriented fair ranking framework that allocates resources on demand to meet the personalized needs of different providers.
Accordingly, we propose a fair ranking method called EquityRank. 
Experimental results on real-world recommender datasets demonstrate that EquityRank not only effectively meets the tailored provider needs but also better enhances the balance between user effectiveness and provider fairness, thus fostering equity-oriented fairness. 
However, considering only exposure and sales still has certain limitations.
Therefore, in the future, we intend to address the following issues based on this study: Firstly, take other types of provider needs into consideration, such as user loyalty. 
Secondly, the platform's utility can be incorporated into the ranking objective to establish a 3-side balance among users, providers, and the ranking system. 